\documentclass[
    aps,
    reprint,
    amsmath,
    amssymb,
    superscriptaddress
]{revtex4-2}

\usepackage{graphicx}%
\usepackage{multirow}%
\usepackage{amsmath,amssymb,amsfonts}%
\usepackage{amsthm}%
\usepackage{mathrsfs}%
\usepackage[title]{appendix}
\usepackage{xcolor}%
\usepackage{textcomp}%
\usepackage{booktabs}%
\usepackage{url}
\usepackage{algorithm}%
\usepackage{algorithmicx}%
\usepackage{algpseudocode}%
\usepackage{listings}%
\usepackage{comment}
\usepackage{hyperref}
\usepackage{changepage} 
\usepackage{setspace}

\begin{document}

\title{\textbf{Quantum-logic detection of rotationally inelastic collisions of single molecules}}

\author{Mikolaj Roguski}
\thanks{These authors contributed equally.}
\affiliation{Department of Chemistry, University of Basel, Klingelbergstrasse 80, Basel, 4056, Switzerland}
\author{Shimoni Patel}
\thanks{These authors contributed equally.}
\affiliation{Department of Chemistry, University of Basel, Klingelbergstrasse 80, Basel, 4056, Switzerland}
\author{Richard Karl}
\affiliation{Department of Chemistry, University of Basel, Klingelbergstrasse 80, Basel, 4056, Switzerland}
\author{Aleksandr Shlykov}
\affiliation{Department of Chemistry, University of Basel, Klingelbergstrasse 80, Basel, 4056, Switzerland}
\author{Octavio Roncero}
\affiliation{Instituto de Fisica Fundamental (IFF-CSIC),C.S.I.C., Serrano 123, Madrid, 28006, Spain}

\author{Markus Meuwly}
\email{m.meuwly@unibas.ch}
\affiliation{Department of Chemistry, University of Basel, Klingelbergstrasse 80, Basel, 4056, Switzerland}

\author{Stefan Willitsch}
\email{stefan.willitsch@unibas.ch}
\affiliation{Department of Chemistry, University of Basel, Klingelbergstrasse 80, Basel, 4056, Switzerland}

\begin{abstract}
Collisions between molecules govern chemical reactions and energy transfer across a range of different environments from interstellar space to the Earth's atmosphere. Despite their fundamental importance, molecular collisions have traditionally been studied through ensemble-averaged experiments which obscure the dynamics of individual scattering events on the quantum level. Probing single molecular collisions with quantum-state resolution has thus far remained a major experimental challenge. Here, we demonstrate a method for detecting individual rotationally inelastic collisions of single molecular ions with neutral atoms using quantum-logic techniques. We prepare individual trapped N$_2^+$ ions in their rotational ground state, subject them to a beam of helium atoms and detect isolated collision events through their effects on the motion of the trapped ions. Quantum-logic-spectroscopic measurements reveal changes of the molecular state upon collision enabling the determination of the probability of rotationally inelastic scattering. The experiments are analysed by comparison with quantum and classical scattering calculations yielding detailed insights into the underlying collision dynamics. By combining single-collision detection with non-destructive readout of the molecular quantum state, our approach establishes a new experimental framework for investigating molecular scattering beyond conventional ensemble measurements. It opens a route towards investigating state-to-state scattering in individual molecules and exploring molecular interactions at the level of single quantum-state resolved events.
\end{abstract}

\maketitle


\section*{Introduction}\label{sec:introduction}

Our understanding of the interactions and chemical reactions of molecules is built around studying collisions between them \cite{levine2009molecular}. The relevant experimental observables, such as collision rates, final-quantum-state distributions and product branching ratios, are typically averages over the parameters of individual collisions including the initial quantum states of the collision partners, their kinetic energies, their impact parameters, and their relative orientations \cite{brouard10a, Pan2017wct, Pan2020abnh}. Such averaging obscures individual dynamic pathways, hampers the characterization of competing processes and obstructs the investigation of fine details of collision processes. Therefore, new types of experiments are required which allow the study of individual molecular-collision events on the quantum level and thus achieve the experimental precision needed to gain detailed insights into the dynamics of molecular scattering processes \cite{balakrishnan2016perspective-544, Bohn20179iu, toscano2020cold-40b, dei2024cold-a3b, Karman2024, Liu2021}.

Precise investigations of collisions of individual particles require both, the detection of isolated scattering events and the control and readout of the internal quantum states of the collision partners. In this context, sophisticated experimental tools conceived within the realms of quantum information and atomic clocks have been developed to probe and manipulate the internal states of single atomic ions confined in radiofrequency traps \cite{leibfried2003quantum-7c4, schmidt2005spectroscopy-2b6}. These techniques have recently also been adapted to studies of collisions of single atoms \cite{katz2022quantum-36d, feldker2020buffer-892, weckesser2021observation-0b4} which yielded detailed insights into the dynamics of their collisions on the quantum level, thus demonstrating the scientific potential of this approach. 

By contrast, molecules are more challenging to control than atoms. Only recently, advances in the quantum manipulation of single molecular ions have enabled the non-destructive readout of their quantum states \cite{wolf2016non-destructive-ecf, chou2017preparation-d64, sinhal2020quantum-nondemolition-757, holzapfel2025quantum-4e2}. In this approach, conventionally dubbed quantum-logic spectroscopy (QLS), the internal state of a trapped molecular ion is mapped onto the state of a co-trapped atomic "logic" ion from which it is subsequently read out. As the molecular state is preserved upon readout, the repeated non-destructive and therefore highly sensitive interrogation of the same molecule becomes possible. 

Building on our previous work~\cite{sinhal2020quantum-nondemolition-757, najafian2020identification-ddb}, we demonstrate here a method based on QLS to probe individual rotationally inelastic collisions of single trapped molecular ions with neutrals thus introducing quantum-logic techniques to molecular-collision studies. Our approach is demonstrated with the prototypical N$_2^+$+He collision system as an example. The collisions were initiated by passing a beam of He atoms through an ion trap containing single state-selected N$_2^+$ ions coupled to single Ca$^+$ atomic ions. Individual scattering events were detected by monitoring the spatial rearrangement of the ions in the trap. The state of the molecular ion was probed before and after each collision using a QLS scheme \cite{sinhal2020quantum-nondemolition-757}. By correlating collisions with state changes, the ratio of elastic to rotationally inelastic collisions was determined. The experimental results were compared with quantum- and classical-scattering calculations performed on an ab-initio potential energy surface yielding insights into the detailed scattering dynamics of the N$_2^+$-He system. 

The present approach is complementary to experiments with neutral molecules trapped in optical tweezers which have recently also reached the level of single particles \cite{liu2018building-bdb, cheuk2020observation-fd0, Vilas2024}. It opens up a new paradigm for studies of molecular collisions in which individual scattering events are probed with high sensitivity on the quantum level using quantum-logic techniques. The current work also introduces a new general approach for studying inelastic processes of molecular ions which are of particular relevance in atmospheric, plasma and interstellar environments, but for which data availability is extremely limited due to the lack of widely applicable experimental techniques so far. The demonstrated methodology paves the way for realizing state-to-state scattering studies on the single-molecule and single-quantum-state level and thus for studying molecular interactions in unprecedented detail.



\section*{Experimental approach}\label{sec:results-experimental-sequence}

\begin{figure}
    \centering
    \includegraphics[width=\columnwidth]{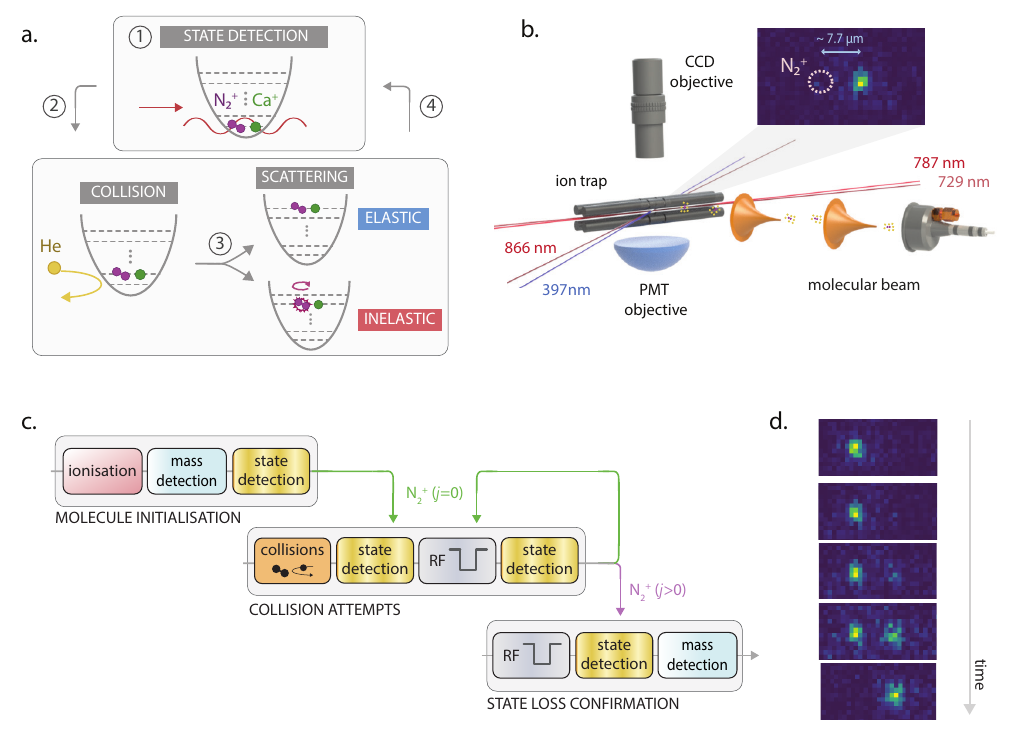}
    \caption{\textbf{Experimental scheme}. \textbf{a.} Experimental sequence for probing rotationally inelastic scattering of single N$_2^+$ ions with He atoms using quantum-logic spectroscopy. Following the preparation of a N$_2^+$-Ca$^+$ two-ion string, the rotational state of N$_2^+$ was measured before and after each passage of a molecular beam containing He (steps \textcircled{1} and \textcircled{4}). Collisions of the N$_2^+$ ions (step \textcircled{2}) resulted in either elastic or inelastic scattering of the molecule (step \textcircled{3}). Transfer of kinetic energy during the collision resulted in motional excitation of the initially stationary ion pair detected as ion motion with a CCD camera and as changes in Ca$^+$ resonance fluorescence measured with a photomultiplier tube (PMT). 
    \textbf{b.} Schematic of the experimental setup consisting of a molecular beam coupled to a linear radiofrequency ion trap. See text for details.
    The inset shows a false-colour fluorescence image of a N$_2^+$-Ca$^+$ two-ion string. Only the resonance fluorescence of the laser-cooled Ca$^+$ ion is visible, the position of the non-fluorescing molecular ion is indicated by the circle. \textbf{c.} Schematic of the experimental sequence. 
    \textbf{d.} Series of images corresponding to an ion reordering event.}
    \label{fig:fig1-experimental-setup}
\end{figure}

The present experiments relied on the initial state preparation of single N$_2^+$ ions in their rotational ground state \cite{shlykov2023optimized-b81,roguski2025role-43b} followed by monitoring the preservation or change of the state upon detection of single collision events with He atoms from the molecular beam. The salient experimental steps are illustrated in Fig. \ref{fig:fig1-experimental-setup}a. The setup was based on a linear radiofrequency ion trap coupled to a molecular-beam machine \cite{karl2026automation-244}, as illustrated schematically in Fig. \ref{fig:fig1-experimental-setup}b. Details are described in the Methods. The experimental protocol consisted of three steps: molecular-state initialisation, collision attempts and state-loss confirmation, as shown in Fig. \ref{fig:fig1-experimental-setup}c. An experimental cycle was initiated with the preparation of a laser-cooled N$_2^+$–Ca$^+$ two-ion string (inset in Fig.~\ref{fig:fig1-experimental-setup}b) \cite{meir2019state-selective-29e, karl2026automation-244}. The molecular ion was prepared in its rotational ground state using state-selective resonance-enhanced multiphoton ionization of a neutral N$_2$ molecule from the molecular beam \cite{shlykov2023optimized-b81} followed by a verification of the rotational state with QLS. The QLS-state-interrogation protocol was based on an optical-dipole force dependent on the rotational state of N$_2^+$ which was set to cause coherent motional excitation of the ion string if the molecular ion was in its rotational ground state. The motion of the ions was then detected by Rabi-sideband thermometry on the Ca$^+$ ion (see Methods and Refs. \cite{sinhal2020quantum-nondemolition-757, roguski2025role-43b} for details). The Ca$^+$ ion served several roles in the present experiments: as a coolant, as a probe for collision detection and as "logic" ion for quantum-logic state detection.   

Collisions were initiated by passing a pulsed molecular beam containing a mixture of He and N$_2$ with a ratio of 7:1 through the ion trap. The number of gas pulses was chosen such that the overall observed collision probability was less than 50\% per experimental cycle (Methods). The small fraction of N$_2$ in the beam was required for the initial generation of the N$_2^+$ ions as outlined above.

After each cycle of exposure of the molecular beam to the ions, the rotational state of N$_2^+$ was probed. In addition, the mass of the molecular ion was determined via resonant motional excitation \cite{drewsen2004nondestructive-8a8} to discriminate between genuine state-changing collisions and spurious chemical reactions of the trapped N$_2^+$ with background gas.

\section*{Observation of single collisions}\label{sec:results-observation-of-single-collisions}

\begin{figure}[!h]
    \centering
    \includegraphics[width=0.95\linewidth]{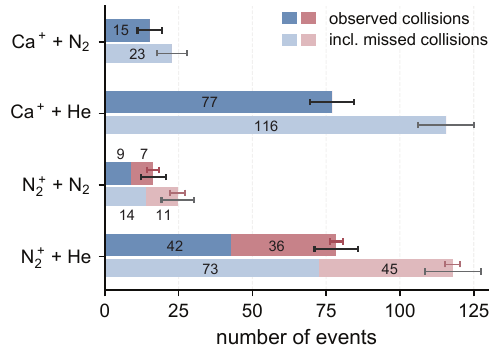}
    \caption{\textbf{Collision statistics.} Number of collisions for different collision pairs when considering only observed collisions (dark bars) and additionally including missed reordering events (light bars). The expected proportion of rotationally inelastic collisions with the trapped N$_2^+$ is shown in red, and elastic collisions causing no rotational excitation in blue. The dark and light-grey error bars refer to the uncertainties of the total number of collisions for each pair, and account for errors arising from multinomial sampling, collision misidentification, and uncertainties in the exact proportions of individual gas components in the molecular beam (see Secs. \ref{sec:unc-in-collsion-event-observation} and \ref{sec:uncertainty-of-N2-state-change} of the SI for details). The dark and light-red error bars refer to the uncertainties in the number of expected state changes for the collision pairs with N$_2^+$.}
    \label{fig:fig3-collision-statistics}
\end{figure}

Individual scattering events were detected by monitoring the effect of kinetic energy imparted to the trapped ions upon collisions. Collisional heating of the ions could cause a swap in the position of the two ions in the trap (reordering) after the kinetic energy was dissipated again via laser cooling, or affect the Ca$^+$ fluorescence signal via the Doppler effect \cite{hankin2019systematic-82e}. Changes in the position of the N$_2^+$ and Ca$^+$ ions on the camera (Fig.~\ref{fig:fig1-experimental-setup}d) were used as a primary marker for identifying single collision events, besides temporary changes of the resonance fluorescence of Ca$^+$ and decrystallisation of the ion string which were also accounted for (see Sec.~\ref{sec:collision-statistics} in the SI for a detailed discussion of the analysis). A similar approach has previously been employed in the determination of collision rates of trapped atomic ions with residual background gas in the vacuum \cite{pagano2018cryogenic-b58,spivey2021high-stability-f40, hankin2019systematic-82e, aikyo2020vacuum-e20}. Following Ref.~\cite{hankin2019systematic-82e}, a position exchange occurs with a statistical probability of close to 50\% when the kinetic energy of the ions exceeds their reordering energy in the trap. The remaining fraction remained undetected and must be accounted for in the analysis. For the trap parameters used in this study, this reordering barrier was evaluated to be $\approx$0.1~meV \cite{hankin2019systematic-82e}.

From a total of 187 collision events observed in the present experiments, 43 resulted in loss of the rotational ground state and thus rotational excitation of N$_2^+$. In addition, 94 missed expected reordering events and 13 detected state changes that occurred during the molecular beam exposure but were not accompanied by any collision marker needed to be included in the analysis (Sec.~\ref{sec:undetected-collisions} of the SI). Collisions could occur between either of the two trapped ions (N$_2^+$, Ca$^+$) and either of the two neutral species (He, N$_2$) present in the beam. To determine the number of collisions corresponding to the system of present interest, i.e., N$_2^+$ and He, it was assumed that the probability of collisions corresponding to each of the four collision-partner pairs was proportional to the relevant Langevin rate constants \cite{Gioumousis1958} taking into account the 1:7 mixing ratios of N$_2$ and He in the molecular beam (see Sec.~\ref{sec:collision-partner-assignment} in the SI for details). Fig. \ref{fig:fig3-collision-statistics} shows the number of elastic (blue bars) and inelastic (red bars) collisions for each collision pair evaluated when considering only observed events (dark bars) and all events including the missed ones (light bars). 

\section*{Inelastic scattering probability}\label{sec:discussions-inelastic-scattering-probability}
Collisions resulting in rotational excitation of N$_2^+$ could occur with either of the two neutral species in the molecular beam, i.e., He or N$_2$. To estimate the probability of inelastic scattering, and therefore the number of inelastic scattering events for the N$_2^+$+N$_2$ pair, we performed a series of control measurements with a molecular beam containing only N$_2$ (Sec.~\ref{sec:appendix-experiemnts-with-pure-N2-beam} of the SI). The expected number of N$_2^+$+He collisions was then obtained by subtracting the expected number of N$_2^+$+N$_2$ collisions from the total number of observed inelastic collisions. The resulting number of rotationally inelastic collisions for the collision pairs when considering all observed events and including missed ones is indicated as the dark and light red bars, respectively, in Fig. \ref{fig:fig3-collision-statistics}.

Having estimated the expected number of collisions between N$_2^+$ and He, and the number of inelastic collisions that correspond to them, a probability of 45(5)\% for inelastic scattering of N$_2^+$ was determined for the observed sample and of 38(4)\% when taking into account the missed reordering events (Sec.~\ref{sec:extracting-ratio-of-elastic-to-inelastic} of the SI).

\section*{Comparison with theoretical calculations}

\begin{figure*}[t]
\centering
\includegraphics[width=.8\textwidth]{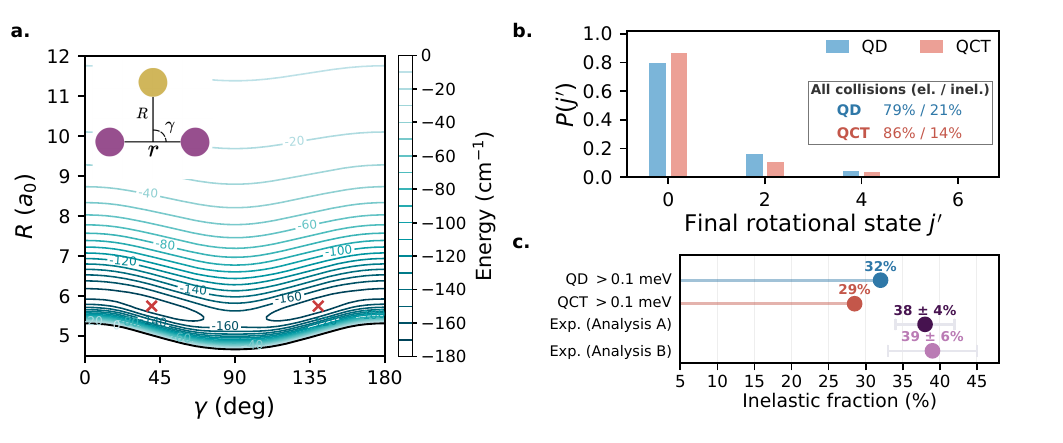}
\caption{
\textbf{Potential energy surface, final-state distributions and inelastic scattering probabilities for N$_2^+$–He.}
\textbf{a.} Potential energy surface of the $\mathrm{N_2^+}({}^2\Sigma_g^+)$--He system calculated at the $\mathrm{N_2^+}$ equilibrium bond length of $r = 2.12\,a_0$. The Jacobi coordinates used to describe the system are illustrated in the inset. The red crosses indicate the symmetry-equivalent minima corresponding to the equilibrium geometries of the $\mathrm{N_2^+}$--He complex.
\textbf{b.} Distribution of final rotational states $P(j')$ for collisions
$\mathrm{N_2^+}(v=0,j=0) + \mathrm{He}
\rightarrow
\mathrm{N_2^+}(v'=0,j') + \mathrm{He}$
at an energy of $33.4\,\mathrm{meV}$, obtained from quantum-dynamics (QD) calculations and from quasi-classical trajectory (QCT) calculations. \textbf{c.} Comparison of elastic and inelastic scattering probabilities obtained from QD and QCT calculations as well as the experiment considering only collisions that transfer more than 0.1 meV of kinetic energy to N$_2^+$ in the laboratory frame. The experimental values were calculated accounting for statistical uncertainties only (Analysis A), and for additional systematic effects due to the collision model assumed in the analysis of the experiments (Analysis B). The error bars indicate $1\sigma$ uncertainties. See text for details.
}
 \label{fig:Figure3}
\end{figure*}

To gain insights into the detailed collision dynamics between N$_2^+$ and He, quantum dynamics (QD) and quasi-classical trajectory (QCT) calculations were performed on a full-dimensional ab-initio potential energy surface of the collision system (Methods). These calculations yielded the distribution of N$_2^+$ rotational states after a collision and the relative proportion of elastic and rotationally inelastic scattering channels for comparison with experiments. A representative contour plot of the potential energy surface (PES) of the system with the N$_2^+$ bond length fixed at its equilibrium value of $r=2.12~a_0$ is shown in panel a of Figure~\ref{fig:Figure3}. 

Owing to the small polarisability of He, the interaction with N$_2^+$ is weak leading to a flat PES with two shallow, symmetry-equivalent minima with a computed well depth of $-175$~cm$^{-1}$ (21.7 meV) at an N$_2^+$-He distance of $R=5.7~a_0$ and Jacobi angles of $\gamma=40^\circ$ and $140^\circ$. The anisotropy of the PES visible in panel a of Fig.~\ref{fig:Figure3} is responsible for rotationally inelastic scattering of N$_2^+$.

The probabilities for scattering into a final rotational state of N$_2^+$ with rotational quantum number $j'$ from the rotational ground state $j=0$ are shown in Fig. \ref{fig:Figure3}b. The collision energy was set to 33.4 meV in the calculations corresponding to the average value in the experiments. Note that collisions with He are not expected to change the nuclear-spin symmetry of N$_2^+$ \cite{Quack1977}. Therefore, only even-$j'$ rotational levels can be populated starting from $j=0$.

The QD and QCT calculations generally give similar results for the distribution of final rotational states. The QD calculations (blue bars) predict that elastic scattering constitutes $79\%$ of all events, while inelastic processes contribute $21\%$ and that the dynamics are dominated by scattering into low-$j'$ channels. After accounting for nuclear-spin statistics by restricting histogram binning to the allowed rotational states (Sec.~\ref{sec:computational_details} of the SI), the QCT calculations predict an elastic contribution of $86\%$. The relevant absolute elastic and inelastic cross sections are reported in Tab.~\ref{tab:elastic_inelastic_cross_sections} of the SI.  

However, it needs to be taken into account that most elastic collisions are glancing at large impact parameters and impart too little kinetic energy to the ion to cause ion reordering. As discussed above, these events cannot be detected in the experiments. Taking into account only the fraction of elastic collisions that transfer more than 0.1~meV of kinetic energy which corresponds to the height of the barrier for reordering, the fraction of inelastic to detectable elastic collisions is computed as 32\% and 29\% for the QD and QCT treatments, respectively (Fig.~\ref{fig:Figure3}c).

\begin{figure*}[t]
\centering
\includegraphics[width=0.8\textwidth]{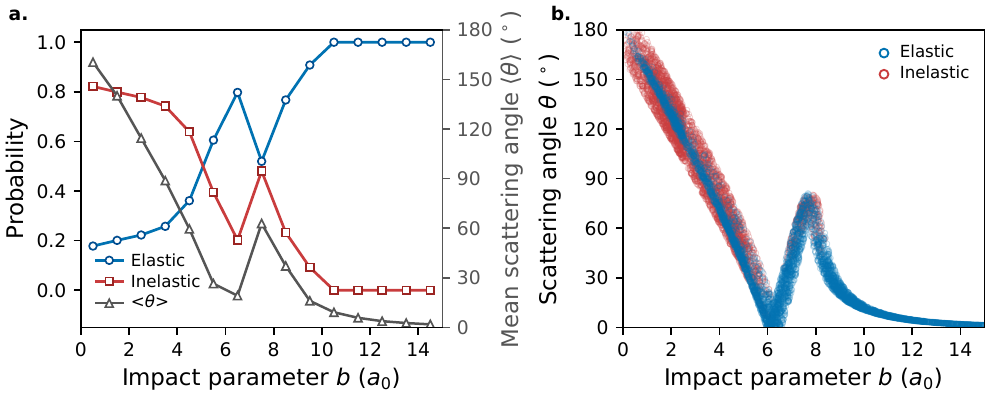}
\caption{\textbf{Opacity and deflection functions for N$_2^+$–He scattering.}
\textbf{a.} QCT elastic and inelastic scattering probabilities as a function of impact parameter $b$, together with the mean scattering angle $\langle\theta\rangle$. The local enhancement of the inelastic probability near $b \approx 7.7$ $a_0$ coincides with a change in the mean scattering angle consistent with a rainbow-scattering feature. \textbf{b.} Deflection function for elastic and inelastic QCT trajectories.}
 \label{fig:Figure4}
\end{figure*}

Qualitative insight into the scattering dynamics of N$_2^+$+He can be gained from inspection of the classical opacity and deflection functions displayed in Figs.~\ref{fig:Figure4}a and b, respectively. Both figures show that inelastic scattering events (indicated by red lines and symbols) dominate at small impact parameters $b$ and, therefore, small values of the total-angular-momentum quantum number $J$. This is also evident from the quantum analogon of Fig.~\ref{fig:Figure4}a shown in Fig.~\ref{fig:FigureS5} of the SI. This indicates that collisions which sample the pronounced anisotropy of the potential energy surface at short distances between the collision partners (see Fig.~\ref{fig:Figure3}) promote changes of the rotational angular momentum. From Fig.~\ref{fig:Figure4}b, it is evident that these short-range inelastic collisions also entail large scattering angles. Elastic processes (blue lines and symbols), by contrast, are dominant at large impact parameters, where the anisotropy of the PES is weak, and are mostly glancing, i.e., forward scattered.  

Two details of the dynamics are noteworthy. First, note the characteristic revival of the inelastic scattering probability as a function of impact parameter which traverses through a second maximum around $b=7.7~a_0$ in both the opacity and deflection functions in Figs.~\ref{fig:Figure4}a and b. Following Refs.~\cite{morita2021,wang2026}, we tentatively attribute this maximum to inelastic rainbow scattering, i.e., the confluence of different scattering paths at the same scattering angle for impact parameters close to the region of maximum attraction in the potential (here $\sim 6~a_0$). Second, Fig.~\ref{fig:FigureS6} of the SI shows that collisions close to a T-shaped approach geometry, i.e., at Jacobi angles $\gamma\approx 90^\circ$, are elastic even at low impact parameters. These collisions hit N$_2^+$ sideways, are predominantly backward scattered in the process and thus exert negligible torque on the molecule so that no rotational excitation ensues. By contrast, collisions at approach angles deviating from 90$^\circ$ lead to increasing rotational excitation due to the larger torque imparted on the molecule.

\section*{Discussion}

While experiment and theory for the inelastic scattering fractions are in broad agreement (29 and 32\% for the QCT and QD calculations, respectively, vs. 38(4)\% for the experiment, see Fig.~\ref{fig:Figure3}c), it is worth exploring possible causes for the remaining discrepancy in more detail. For the experiments, a detailed analysis of the presented uncertainties is given in Secs.~\ref{sec:unc-in-collsion-event-observation} and \ref{sec:uncertainty-of-N2-state-change} of the SI. A potentially significant systematic uncertainty in the analysis of the experimental data which cannot be accurately quantified at the moment lies in the application of the Langevin capture model to estimate the expected number of collisions for each pair of collision partners. While Langevin theory, which considers only spiralling collisions, may be an adequate approximation for treating close-range inelastic events, this approach underestimates the cross sections of long-range elastic processes. 

To assess the robustness of the present experimental value of the inelastic collision fraction with respect to the assumed collision model, a semi-classical theory for elastic collisions similar to ones previously used in the literature \cite{hankin2019systematic-82e} was employed as an alternative methodology to gauge the contributions of the different collision partners to the observed scattering events. This approach yielded the same results as the Langevin treatment within the experimental error limits (Sec.~\ref{sec:extracting-ratio-of-elastic-to-inelastic} of the SI). Another factor impacting the analysis of the experimental data could be the higher mass of N$_2$ molecules and thus their higher collision energy in comparison to that of He atoms in the mixed beam, as well as in comparison to the pure N$_2$ beam used for the control measurements (see Tab.~\ref{tab:collisional-energies} of the SI). These differences may affect the probability of observing N$_2^+$+N$_2$ over N$_2^+$+He collisions, in particular glancing elastic collisions with an energy transfer close to the reordering threshold.

To estimate the possible ramifications of these systematic effects, a sensitivity analysis was performed (see Sec.~\ref{sec:systematic-errors-in-data-analysis} of the SI for details). Here, the inelastic scattering probability for N$_2^+$+He was recalculated using Monte-Carlo sampling over a range of rate coefficients for each collision pair which were allowed to deviate by $\pm30$\% from the relevant Langevin value. Additionally, the inelastic scattering probability for N$_2^+$+N$_2$ was allowed to vary by up to $\pm 30$\%. From this procedure, the mean state-change probability for N$_2^+$+He (when including missed reorderings) was determined to be ($39 \pm 3_{\mathrm{stat.}} \pm 5_{\mathrm{sys.}}$)\%, and ($39 \pm 6$)\% when including both statistical and systematic uncertainties in the sampling procedure (magenta data point in Fig.~\ref{fig:Figure3}c). This result agrees with the nominal value of ($38\pm4$)\% within the error boundaries.

On the theory side, although the inelastic fraction from QCT and QD simulations agrees to within 10 \% (29 \% vs. 32 \%, see Fig. \ref{fig:fig3-collision-statistics}), the computed cross sections can differ by up to 30 \% (Table \ref{tab:elastic_inelastic_cross_sections}). In addition, the computed PES may be further improved. 

Detailed QD calculations on accurate potential surfaces for the three remaining pairs of collision partners in the present experiments in addition to N$_2^+$+He would be required to obtain more accurate theoretical estimates of all relevant collision cross sections. On the experimental side, introducing separate molecular beams for N$_2$ and He would enable a separation of the different collision pairs in the scattering experiments thus simplifying the analysis of the experimental data. Such calculations and developments are an objective for future studies.

\section*{Conclusions and outlook} \label{sec:conclusions}

The present work combined quantum-logic with chemical-physics techniques to study individual molecular scattering events with rotational-quantum-state resolution. The present methodology benefits from a synergy of two components: detection of single molecular collisions, which eliminates ensemble averaging, and quantum-logic spectroscopy for non-destructive molecular-state readout, which enables unprecedented levels of detection sensitivity in collision experiments. As demonstrated here, the current method is particularly suited for the study of inelastic collisions and addresses the scarcity of state-resolved experimental techniques for investigations of such processes in molecular ions. 

In the present work, the identification of molecular state was limited to monitoring rotational excitations from the ground state. However, the current quantum-logic approach can also be extended to investigate higher rotational states directly as demonstrated in Refs.~\cite{najafian2020identification-ddb, roguski2025role-43b} and it can be combined with laser-based protocols that allow state preparation of single molecular ions on the hyperfine-Zeeman level \cite{shlykov2026quantum}. Incorporating these elements into the present experiment will allow the probing of scattering events on the level of fine, hyperfine and even Zeeman states \cite{liu2025hyperfine-to-rotational-84b} and enable the detection of both the initial and final states of the collision process. This paves the way for studies of state-to-state scattering of single molecules on the level of individual quantum states which has been a long-standing objective in the realm of molecular and chemical physics. 

The presented technique is flexible in its choice of collision partners, and can be applied to study other systems by suitable adaption of the quantum-logic protocol \cite{wolf2016non-destructive-ecf, chou2017preparation-d64, holzapfel2025quantum-4e2}. The sensitivity and state-resolution achieved here provides a promising platform for exploring quantum effects related to ion-neutral collisions, such as Feshbach resonances \cite{weckesser2021observation-0b4, narevicius:2023} and quantum interference \cite{katz2026quantum-eed} demonstrated for atomic ions, or collisional entanglement \cite{wang2025motional-254, devolder2026general-f4d}. Moreover, the sensitivity reached in the present experiments paves the way towards probing individual collision channels and potentially even steric effects \cite{paliwal2026control-d2e} and for highly precise validations of molecular potential-energy surfaces and scattering calculations \cite{horn:2025}.


\section*{Methods}\label{sec:methods}

\subsection*{Experimental procedures}
The experimental setup consisted of a linear radiofrequency ion trap coupled to a molecular-beam machine as schematically shown in Fig.~\ref{fig:fig1-experimental-setup} and detailed in Refs. \cite{meir2019state-selective-29e, sinhal2020quantum-nondemolition-757, karl2026automation-244}. Each experimental cycle started with loading Ca$^+$ and N$_2^+$ ions into the trap. The ion string was continuously laser cooled to mK temperatures on the $(4s)~^2S_{1/2}\leftrightarrow (4p)~^2P_{1/2}\leftrightarrow(3p)~^2D_{3/2}$ optical-cycling transitions of Ca$^+$ during all experimental stages. The presence of a dark N$_2^+$ ion was confirmed by mass spectrometry based on resonant motional excitation of the two-ion chain. The preparation of the ground rotational state of the molecular ion with quantum number $j$=0 was confirmed using QLS before passing the molecular beam through the trap and inducing collisions. After each cycle of passing the molecular beam through the trap, the state of N$_2^+$ was checked at least twice. Ramping down and subsequently restoring the amplitude of the trap RF field between sets of state-detection pulses ('RF' in Fig.~\ref{fig:fig1-experimental-setup}c) helped to re-crystallize the ions in case a decrystallisation event occurred. If state loss was detected, the mass of the molecular ion was measured to distinguish between inelastic and reactive collisions. 

The quantum-non-demolition state-detection protocol used for QLS of N$_2^+$ was based on an optical dipole force dependent on the molecular state that resonantly excited the axial in-phase motional mode of the N$_2^+$-Ca$^+$ string in the trap \cite{sinhal2020quantum-nondemolition-757}. Prior to QLS interrogation, this mode was cooled close to its motional ground state by resolved-sideband techniques \cite{roguski2025role-43b}. A travelling optical lattice was then applied to exert the state-dependent optical dipole force which caused coherent motional excitation of the two-ion crystal only when N$_2^+$ was in its rotational ground state. The motional excitation was read out via sideband thermometry on a Zeeman component of the $(4s)~^2S_{1/2}\rightarrow (3d)~^2D_{5/2}$ clock transition of Ca$^+$ around 729~nm. Typically, 40 QLS measurements were performed per state-detection attempt corresponding to a state-determination fidelity $>$99.9\% \cite{roguski2025role-43b}. 

The same molecular beam was used in the present experiments as a source of neutral N$_2$ for photoionization and of He atoms for collisions. A mixing ratio N$_2$:He of 1:7 in the beam was prepared by combining He (purity grade 6.0) and N$_2$ (purity grade 7.0) gas in a lecture bottle feeding a pulsed gas valve. The backing pressure of the valve was kept at 2 bars. The gas mixture was expanded through the valve at a repetition rate of 10 Hz at an opening time of $\approx$220 $\mu$s. 

Single N$_2^+$ ions in their rotational ground state were produced by [2+1'] resonance-enhanced photoionisation of N$_2$ from the beam followed by verification of their quantum state using QLS \cite{shlykov2023optimized-b81}. For the collision experiments, the trapped N$_2^+$ ions were exposed to the molecular beam for a duration of 2 seconds (21 gas pulses) per experimental cycle totalling $\approx$4.6 ms of effective contact time between the beam and the ions. The mean longitudinal velocity of the molecular beam was determined to be $\sim$1350~ms$^{-1}$ defining the collision energy between N$_2^+$ and He of 33~meV (Sec.~ \ref{sec:collision-energies} of the SI).

\subsection*{Computations}\label{sec:theoretical-studies}

\subsubsection*{Potential energy surface}

The interaction between N$_2^+$ and He was described using a three-dimensional potential energy surface (PES) calculated \textit{ab initio} using the unrestricted coupled-cluster method with single, double, and perturbative triple excitations (UCCSD(T) \cite{ccsdt}) method in conjunction with the aug-cc-pVQZ basis set \cite{basisset,basisset1} as implemented in the \textsc{Molpro} software package \cite{molpro}. 

The PES was constructed in Jacobi coordinates $(R, r, \gamma)$, where $R$ is the distance between the He atom and the centre of mass of N$_2^+$, $r$ is the N--N bond length, and $\gamma$ is the angle between the molecular axis and the incoming He atom (inset in Fig.~\ref{fig:Figure3}a). The intermolecular distance $R$ was sampled using 50 points in the range 2.0--10.0~\AA, with a spacing of $\sim$0.1~\AA\ in the short-range region (2.0--5.0~\AA) and $\sim$0.25~\AA\ at larger separations (5.0--10.0~\AA). The diatomic distance $r$ was sampled using 11 points between 0.998~\AA\ and 1.288~\AA. 
The angular coordinate $\gamma$ was represented using an 11-point Gauss--Legendre quadrature. Owing to the symmetry of
the homonuclear N$_2^+$ molecule, the potential satisfies $E(r,R,\gamma)=E(r,R,180^\circ-\gamma)$. The central grid point
was evaluated at $\gamma=89^\circ$ rather than exactly $90^\circ$ so that all electronic-structure calculations could
be performed consistently in the $C_s$ point group. Thus, a total of $M = 6600$ reference energies was obtained. The origin of the energy scale was set to the asymptotic limit of the separated N$_2^+$ and He fragments with N$_2^+$ in its optimized geometry.

The PES was represented as a reproducing kernel Hilbert space (RKHS) \cite{rkhs} by expressing it as a weighted sum over all \textit{ab initio} data points,
\begin{equation}
    V(x) = \sum_{k=1}^{M} \alpha_k \, Q(x_k, x),
\end{equation}
where $M$ is the total number of \textit{ab initio} geometries, and $x = (R, r, z)$ represents the set of internal coordinates. The angular coordinate is transformed as $z = (1 - \cos\gamma)/2$, while $x_k = (R_k, r_k, z_k)$ corresponds to a grid point. The kernel function $Q(x_k, x)$ was a product of one-dimensional kernels along each coordinate,
\begin{equation}
    Q(x_k, x) = q^{2,4}(R_k, R)\, q^{2,4}(r_k, r)\, q^{2}(z_k, z),
\end{equation}
where the functions $q^{2,4}$ and $q^{2}$ describe the radial and angular dependences of the interaction, respectively \cite{rkhs}.

The quality of the RKHS PES was evaluated using a correlational plot comparing \textit{ab initio} and RKHS energies as shown in Figure~\ref{fig:FigureS4} of the SI. The mean absolute deviation is 0.09~cm$^{-1}$ for the on-grid points and 1.06~cm$^{-1}$ for the off-grid points which demonstrates the high accuracy of the RKHS-PES.

\subsubsection*{Collision dynamics}

Collisions between N$_2^+$+He were investigated theoretically using two complementary approaches: quasi-classical trajectory (QCT) simulations and quantum dynamics (QD), both performed on the global RKHS potential energy surface.

The present QCT approach followed procedures described in literature \cite{truhlar1979,karplus1965,koner2016,koner2018}. The classical equations of motion derived from the Hamiltonian were integrated using a fourth-order Runge--Kutta algorithm. A time step of $\Delta t = 0.05$~fs was employed ensuring conservation of total energy and angular momentum to high numerical accuracy. The collision energy was fixed at the experimental value of $E_{\mathrm{coll}} = 33.4$ meV and N$_2^+$ was initialized in its rovibrational ground state $j=0$ as in the experiments. Initial conditions for the collisions were generated by Monte Carlo sampling over molecular orientations and impact parameters. Impact parameters were selected using a stratified sampling approach and restricted to a maximum value of $b{_{\max}} = 15~a_0$. In total, $N_{\mathrm{traj}} = 100000$ trajectories were propagated. Since QCT simulations yield classical and, therefore, continuous distribution of post-collision rotational angular momenta, the final rotational states were assigned using histogram binning. The continuous rotational angular momenta were projected directly onto the symmetry-allowed even rotational states using bins centred on the corresponding quantum numbers. The resulting state-resolved populations were renormalised such that the total probability over the included rotational manifold were unity. As an internal consistency check, the same analysis was repeated using Gaussian binning \cite{BONNET1997183} over the identical set of allowed even-$j'$ states, yielding an essentially indistinguishable final rotational state distribution, which is consistent with earlier findings \cite{koner2018}.

The quantum dynamical calculations were performed using the MADWAVE3 program \cite{madwave3}. Here, the wave packet propagation was carried out using a modified Chebyshev propagator. The Hamiltonian was represented in reactant Jacobi coordinates within a body-fixed frame, where the internal degrees of freedom were described by the radial coordinates $R$ and $r$, and the angle $\gamma$ between them. The total wave function was further characterized by the total-angular-momentum quantum number $J$ and its projections $M$ and $\Omega$ onto the space-fixed and body-fixed $z$ axes, respectively. The QD calculations were converged using a basis including total angular momenta up to $J_{\max}=46$.

Further information on the theoretical methodology is given in Sec.~\ref{sec:computational_details} of the SI.


\bibliographystyle{apsrev4-2}
\bibliography{references}



\section*{Acknowledgments}
We thank Georg Holderried, Dominik Baumgartner, Dr. Anatoly Johnson, Grischa Martin and Philipp Knöpfel for technical support.
This work was supported by the Swiss National Science Foundation, grant nrs. 200021\_204123 (SW), TMAG-2\_209193 (SW), 200020\_219779 (MM), 200021\_215088 (MM),
the National Competence Center for Research "Precision", the European Partnership on Metrology (Funder ID: 10.13039/100019599, grant nr. 23FUN04 COMOMET) and the University of Basel (SW and MM). 

\section*{Author contributions}

M.R. performed the experiments and analysed the experimental data. S.P. performed the theoretical calculations with help from O.R. for the QD simulations. R.K. and A.S. assisted in the experiments. S.W. conceived the project. O.R., M.M. and S.W. supervised the project. M.R., S.P. and S.W. drafted the manuscript, all authors contributed to the writing.

\section*{Competing interests}
The authors declare no competing interests.



\clearpage\pagebreak

\renewcommand{\thesection}{S\arabic{section}}
\setcounter{section}{0}
\renewcommand{\thefigure}{S\arabic{figure}}
\setcounter{figure}{0}
\renewcommand{\thetable}{S\arabic{table}}
\setcounter{table}{0}
\renewcommand{\theequation}{S\arabic{equation}}
\setcounter{equation}{0}


\onecolumngrid

\begin{center}
\begin{minipage}{0.82\textwidth}
\begin{center}
\vspace{1.5em}
{\LARGE\bfseries SUPPLEMENTARY INFORMATION\par}




\end{center}
\end{minipage}
\end{center}
\vspace{3em}

\twocolumngrid

\section{Calculation of collision statistics} 
\label{sec:collision-statistics}

\subsection{Observed collisions}

Apart from ion reordering, which was used as the primary marker for identifying individual scattering events discussed in the main text, collisions with the trapped ions were also detected by other experimental observables. 

First, in some cases the collisional energy transferred to the ions was sufficient to temporarily melt the two-ion crystal or to permanently eject one of the ions from the trap. If a temporary disappearance of one of the ions was followed by the retrieval of a two-ion string within a state-detection cycle, a "decrystallisation" event was recorded, and the experimental sequence was continued. Otherwise, the ions were deemed ejected from the trap, and the experiment was reinitialized. Decrystallisation events in which the ions were recovered and the state of N$_2^+$ could be determined were included in the count of observed collision events. Unlike ion reordering, decrystallisation events were considered to be always observable given the large amount of energy transferred to the ions. They did not contribute to the count of missed reordering events (see Sec.~\ref{sec:undetected-collisions} below). 

Second, in some cases collisions could also be detected by monitoring changes of the Ca$^+$ resonance fluorescence within a 1-ms photon binning time during the passage of the molecular beam through the trap. Collisional momentum transfer could affect the fluorescence intensity via the Doppler effect, which shifted the effective laser frequency observed by the ion. Such fluorescence modulation, however, was observed less often than ion reordering, indicating that this method had lower sensitivity for collision detection. This can be rationalized by the fact that the fluorescence intensity was sensitive only to the motional excitation along the direction of the interrogation laser beam, and, more importantly, that the translational energy of the ions was quickly dissipated by the cooling forces. Therefore, the detection efficiency of this method was limited by the detector binning time and background noise.

From the gathered statistics, only 23\% of all ion reordering and decrystallisation events were accompanied by fluorescence loss, while 7 out of 49 collisions identified by fluorescence loss did not exhibit other collision markers. The latter observation confirms the occurrence of missed collisions that should be accounted for. A complete statistics of the recorded collision markers and their correlation with state changes of the trapped N$_2^+$ ions is presented in Table \ref{tab:appendix-col-det-stat}.

Here, the state-change probability of N$_2^+$ of 55 \% observed in decrystallisation collision stands out in the collected statistics. Not only was it much higher than the state-change probability observed during ion reordering, but it was also larger than the $\approx50$\% probability of either of the ions participating in a collision event (compare with Table \ref{tab:langevin-rates-probability}). From the analysis of the camera images, it became apparent that the decrystallisation events with ions subsequently recovered in the trap were largely (65\% probability) due to collisions with N$_2^+$. In these experiments, Ca$^+$ remained visible on the camera but with its position shifted to the trap centre indicating that the molecular ion was ejected to a high orbit in the trap before being sympathetically recooled. The corresponding collisions with the neutral partner were therefore energetic implying small impact parameters and, consequently, an increased probability for inelastic scattering (see Fig.~\ref{fig:Figure3} of the main text). Moreover, as the collision energies between N$_2^+$ and N$_2$ are generally larger than with N$_2^+$ and He (see Tab.~\ref{tab:collisional-energies}), there may have been a bias for observing decrystallisation preferentially for N$_2^+$+N$_2$ collisions.

Furthermore, it is instructive to compare the N$_2^+$ state-change probability associated with ion-reordering events: 15\% in the experimentally detected reordering events and 14\% in the undetected ones. The latter was calculated from a ratio of 13 recorded state changes without any collision markers to the estimated 94 missed ion reorderings. This statistical agreement in inelastic scattering probabilities between observed and missed ion reorderings indicates a common physical origin and supports our treatment of missed events.

\subsection{Treatment of undetected collisions}\label{sec:undetected-collisions}

Collisions could remain undetected for two primary reasons: first, when the kinetic energy imparted to the ions was below the detection threshold, and second, when the collision was statistically unobservable (as for half of the reordering events).

\subsubsection{Collisions below the detection-sensitivity threshold} 
\label{sec:coll_below_threshold}

Collisions that transferred too little energy to the trapped ions to produce any of the experimental observables could not be observed. While inelastic collisions could be detected with high fidelity (Sec.~\ref{sec:uncertainty-of-N2-state-change}), the detection of elastic collisions relied on monitoring the effects of kinetic energy transferred to the ions. 

\begin{figure*}
\centering
\includegraphics[width=0.9\textwidth]{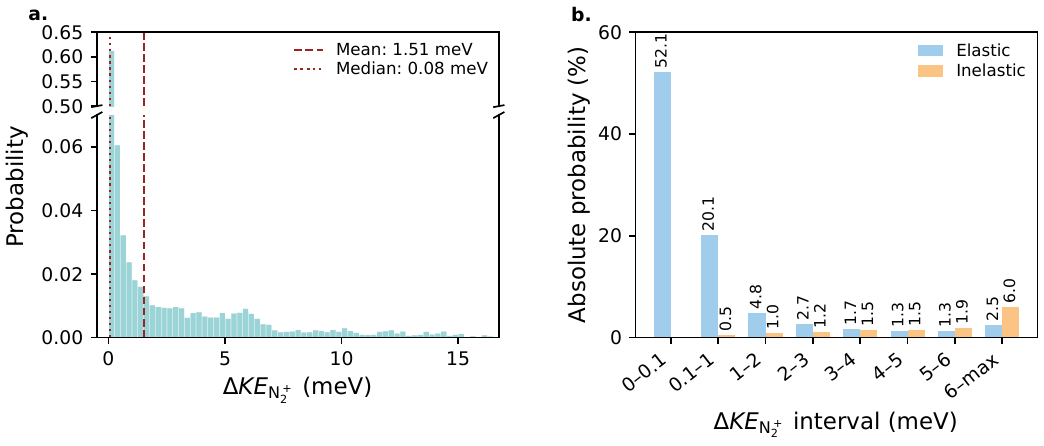}
\caption{\textbf{Kinetic-energy transfer to N$_2^+$ in collisions with He.}
\textbf{a.} Probability distribution of the change in the kinetic energy, $\Delta KE_{\mathrm{N_2^+}}$, of single N$_2^+$ ions after collisions with He atoms obtained from QCT calculations at a collision energy of 33.4 meV. The vertical dashed lines indicate the mean and median values of $\Delta KE_{\mathrm{N_2^+}}$, respectively. The $y$-axis is broken to highlight the high-energy tail of the distribution. 
\textbf{b.} Absolute probabilities of elastic (blue) and inelastic (orange) scattering of N$_2^+$ within different intervals of $\Delta KE_{\mathrm{N_2^+}}$. The probabilities indicated were calculated with respect to the total number of trajectories, such that the elastic and inelastic contributions summed over all $\Delta KE_{\mathrm{N_2^+}}$ intervals represent the complete ensemble of trajectories.}
\label{fig:FigureS1}
\end{figure*}

Figure~\ref{fig:FigureS1}a shows the kinetic energy distribution of N$_2^+$ after collisions with He obtained from the QCT calculations, with a detailed analysis of the energy window between 0 and 6~meV shown in panel b. As discussed in the main text, the elastic collisions were predominantly glancing, occurred at large impact parameters and thus transferred little kinetic energy. Figure~\ref{fig:FigureS1}b shows that about 52\% of the elastic collisions are predicted to transfer less than 0.1~meV corresponding to the energy threshold for ion reordering. These collisions could not be detected in the experiment and need to be accounted for when comparing the experimental results with the theoretical predictions (see main text).

One should also consider factors that could introduce a bias towards certain collision partner pairs being more likely to be involved in collisions that fall below the sensitivity threshold and could, therefore, affect the assignment of collision probabilities discussed in Sec. \ref{sec:collision-partner-assignment} below. One such factor arises from the fact that the energy transferred to the ions in a collision depends on the mass of a neutral collision partner (compare with Eq.~\ref{eq:energy-transfer-ions}). From conservation of momentum and energy, the average kinetic energy imparted to the ions N$_2^+$ and Ca$^+$ is two to three times larger for collisions with N$_2$ than for He for purely elastic collisions. Additionally, the collision energy with N$_2$ is in general four times larger than that with He (see Table \ref{tab:collisional-energies}). Even if part of the collision energy with N$_2$ was partitioned into internal excitation of the neutral partner, collisions with N$_2$ may overall still involve larger kinetic energies transferred to the ions. This may introduce a bias for recording these events because of the longer timescales involved for sympathetic cooling to restore the ion motion to equilibrium. 

Moreover, the efficiency of the dissipation of collision energy depended on the ionic collision partner. In particular, for energetic collisions when one of the ions was temporarily ejected to a high orbit in the trap, cooling of N$_2^+$ required several cooling interactions with Ca$^+$ before crystallisation, whereas Ca$^+$ was directly Doppler-cooled. Thus, collisions with N$_2^+$ were potentially easier to observe because of the longer cooling timescales involved possibly introducing a detection bias. 

The potential ramifications of these effects require further study and were not accounted for in the present analysis.

\subsubsection{Statistically missed ion reorderings}

Once the ions have sufficient energy to overcome the reordering threshold, ion reordering is inherently stochastic. As shown in Ref.~\cite{hankin2019systematic-82e}, the reordering probability quickly but non-linearly increases above the energy threshold to reach approximately 50\%. The energy interval at which the probability converges on 50\% is relatively narrow with an estimated width of $\approx$1~K~=~0.09~meV. Considering the collision energies available in our experiments (see Fig. \ref{fig:FigureS1} and Sec.~\ref{sec:collision-energies}), only a negligible fraction of collision events fall into this specific energy window and would have a reordering probability below 50\%. Therefore, they were omitted in the analysis.

The definition for ion reordering adopted in the present analysis included events in which the ions did not necessarily settle in the exchanged positions, but the reordering dynamics could be observed in the camera images. For 139 reordering events observed in total, one needs to account for 19 recorded events for which the reordering dynamics was sufficiently long so that they could be observed on the camera (at a rate of 3 frames per second), but the ions settled back to their initial positions, as well as for 7 collisions captured only by Ca$^+$ fluorescence-trace analysis. As a result, it is estimated that 139-(2$\times$19)-7=94 collisions went completely undetected. Note that other collision markers such as decrystallisation events were easily identifiable on the camera. Hence, it is assumed that all such events were recorded.

\subsubsection{Total number of collisions}

\begin{figure}
    \centering
    \includegraphics[width=\linewidth]{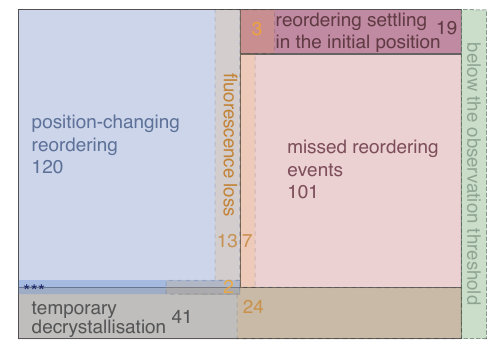}
    \caption{\textbf{Collision statistics.} Breakdown of number of collisions observed by different observables. The primary method of collision identification was based on tracking the positions of the ions in the trap. Two collision markers were observed: ion position exchanges (light blue area) and ion-string decrystallisation (grey area). In three cases, both markers were observed within the same collision attempt (overlapping dark blue area, assigned with '***'), each of the markers corresponding to a separate collision (six collisions in total). Some reordering events were statistically missed (light red). Temporary fluorescence loss of Ca$^+$ (yellow overlapping area) constituted a complementary method for tracking collisions. Some collisions did not provide enough energy to be observed and were missed (green area, not to scale).} 
    \label{fig:collision-statistics-numbers}
\end{figure}

Figure \ref{fig:collision-statistics-numbers} graphically summarises the classification of collision events in our experiments, including the account of undetected collisions. The reordering events could result in either an exchange of positions (120 events, light blue area) or ions settling in their initial positions (19 events, dark red area). Statistically, only 50\% of ion reordering events would result in exchanged positions. Therefore, we inferred the number of missed reordering events (101 events, light red area) from the number of observed reordering events that settled in the exchanged positions, reduced by the observed reordering events that settled in the initial position. Decrystallisation events (41 events, grey area) were distinguished from ion reordering and were considered always observable. The collision attempts associated with two (separate) events of ion reordering and decrystallisation within the same experimental cycle are indicated by the overlapping dark blue area (3+3 events, marked as '***'). The overlaying yellow area corresponds to the Ca$^+$ fluorescence loss events (in total 49 events, partially observed with other collision markers). Lastly, we also acknowledge the proportion of collisions that were not detectable due to insufficient energy transfer (green area, not to scale), as discussed in Sec. \ref{sec:coll_below_threshold}. 


\section{Assignment of collision partners}\label{sec:collision-partner-assignment}

The presence of four possible collision partner pairs introduced an additional layer of complexity in the data treatment. In the current analysis (dubbed model I), it was assumed that the probability of each collision pair participating in a collision event was proportional to their Langevin rate coefficients \cite{Gioumousis1958}, which capture the trends in long-range interactions that dominate the interaction potentials of these systems. These collision probabilities were scaled with the relevant mole fraction of the neutral collision partner (N$_2$ or He) in the molecular beam. The results are listed in Table \ref{tab:langevin-rates-probability}. The quoted uncertainties arise from an imprecision in the pressure readout of $\pm 0.125$~bar during the preparation of the gas mixture of the molecular beam for a final mixing pressure of 8 bar.  

As the Langevin model only takes into account spiralling collisions that reach short range, this approach may be inaccurate for correctly treating long-range glancing elastic collisions \cite{hankin2019systematic-82e}. Thus, for comparison Table \ref{tab:langevin-rates-probability} also lists the probabilities for collision partner-pairs assuming a semiclassical model (dubbed model II) for elastic collision cross sections \cite{ct2000ultracold-f64} similar to the one assumed in the modelling of collisions of Ca$^+$-Al$^+$ two ion crystals with neutral H$_2$ molecules in Ref.~\cite{hankin2019systematic-82e} and in modelling ion transport in molecular gases \cite{yousfi2012basic-5f3, benyoucef2014arar-541}. From \ref{tab:langevin-rates-probability}, it can be seen that the cumulative probability of each trapped ion species (Ca$^+$ or N$_2^+$) to participate in a collision is approximately 50\% regardless of the model used to estimate the collision probabilities. 

The differences in the scattering probabilities of neutral species (He and N$_2$) arising from their differing mole fractions in the beam are further affected by differences in their polarisability and mass. Additionally, the elastic cross sections used in model II exhibit a dependence on the collision energy. This also affects the results because different beam components have different collision energies due to their mass differences (Table~\ref{tab:collisional-energies}). As a result, the probabilities for each collision pair differ between the models. However, the calculation of the final probability for inelastic vs. elastic collisions for N$_2^+$ and He proved insensitive to the collision model assumed and yielded the same result within the error bars (see Sec.~\ref{sec:extracting-ratio-of-elastic-to-inelastic}).


\begin{figure*}
    \centering
    \includegraphics[width=0.95\linewidth]{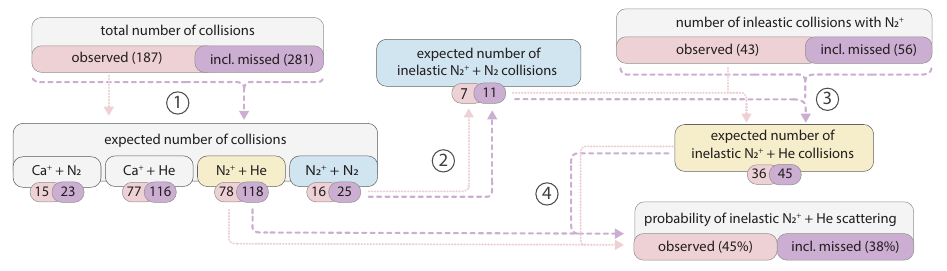}
 \caption{\textbf{Analysis of the experimental results.} Flow chart visualising the steps of the analysis required to determine the probability of inelastic N$_2^+$-He scattering. The two colours of the arrows linking boxes indicate that the analoguous procedure was followed in the treatment of observed collision events and of the missed collisions. See text for details.}
    \label{fig:appendix-analysis-flow}
\end{figure*}

\section{Control experiments with a pure N$_2$ molecular beam}\label{sec:appendix-experiemnts-with-pure-N2-beam}

To obtain an estimate of the inelastic-scattering probability for collisions between N$_2^+$ and N$_2$, a series of control measurements was performed with a molecular beam containing only N$_2$. The experimental procedure was analogous to that used for the experiments with the He/N$_2$ gas mixture discussed in the main text.

The sample consisted of 385 collision attempts, and a breakdown of the observed collision markers is presented in Table \ref{tab:col-det-stat-pureN2}.  The probability of observing a state change per collision event (with either of the two ions in the trap) was 23\%. From a ratio of 52:48 of the Langevin rate coefficients for collisions of neutral N$_2$ with N$_2^+$ and Ca$^+$, it was predicted that 49(5) collision events corresponded to N$_2^+$+N$_2$, and thus the probability of observing a rotational excitation of N$_2^+$ upon detecting a collision with the N$_2$ beam was determined to be 45(5)\%. When accounting for an estimated 70 missed reordering events and including 14 with state change with no associated collision marker, the inelastic scattering probability for N$_2^+$-N$_2$ was found to be 44(3)\%.

Rotational-state change of N$_2^+$ by collisions with N$_2$ can proceed via two distinct mechanisms: inelastic scattering or complex-mediated charge transfer resulting in substitution of the trapped ion with an incoming neutral \cite{tong2012state-selected-d0a}. Details of such interaction do not affect the calculations presented in Sec.~\ref{sec:extracting-ratio-of-elastic-to-inelastic} as only the overall state-changing probability was considered.


\section{Determination of the probability of N$_2^+$ inelastic collisions}\label{sec:extracting-ratio-of-elastic-to-inelastic}

The probability of inelastic collisions of N$_2^+$ with He was calculated in three steps. Fig. \ref{fig:appendix-analysis-flow} summarizes the steps of the analysis. First, the expected number of collisions between different collision partners was obtained by assuming that the probabilities of each pair participating in a collision event are proportional to their relative Langevin rates, as discussed in Sec. \ref{tab:langevin-rates-probability} (\textcircled{1}). The number of inelastic N$_2^+$+N$_2$ collisions was calculated using the inelastic scattering probability determined in control measurements with a pure N$_2$ beam (\textcircled{2}, Sec.~\ref{sec:appendix-experiemnts-with-pure-N2-beam}). For the dataset of observed collisions only, the N$_2^+$-N$_2$ scattering probability was based solely on observed events. When including missed events in the analysis, they were also accounted for in the scattering probability for the experiments for the pure N$_2$ beam. 
This allowed the determination of the expected number of inelastic N$_2^+$-He collisions by subtraction of the inelastic N$_2^+$-N$_2$ instances from all inelastic scattering events (\textcircled{3}). Finally, knowing the expected numbers of inelastic and all collision events between N$_2^+$ and He, the inelastic scattering probability was obtained (\textcircled{4}). The results of this analysis of the collision events were plotted in Figure \ref{fig:fig3-collision-statistics} in the main text. The numerical values are listed in Table \ref{tab:appendix-raw-data-figure-statistics} for reference. 

The experimentally observed state-changing collisions contain contributions from N$_2^+$ colliding with both neutral beam components. The number of inelastic collisions with N$_2$ was estimated using a state change probability of 45(5)\% for N$_2^+$+N$_2$ collisions determined in Sec.~\ref{sec:appendix-experiemnts-with-pure-N2-beam}. Subtracting this contribution from the observed state changes in the experiments with the beam containing both neutrals, the expected number of inelastic N$_2^+$+He collisions was determined to be 36(2). 

This result, however, does not account for the potentially missed collisions discussed in Sec. \ref{sec:undetected-collisions}. Correcting for 13 state changes observed without any collision marker and the expected 94 missed collisions, as well as using the 44(3)\% state change probability per N$_2^+$+N$_2$ (calculated including missed collisions in the control experiments), the probability of inelastic scattering was obtained as 38(4)\%. The results including missed collisions are summarised in Table~\ref{tab:appendix-raw-data-figure-statistics}.


\vspace{2em}
\section{Computational details} \label{sec:computational_details}

\subsection{Potential energy surface}

\begin{figure}
\centering
\includegraphics[width=0.85\columnwidth]{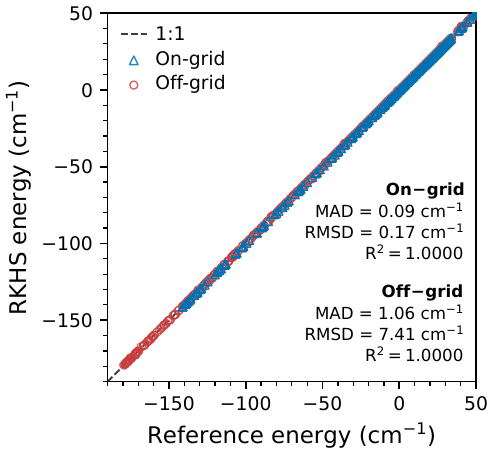}
\caption{\textbf{Comparison between reference {\it ab initio} energies and the RKHS-PES} for on-grid (blue) and off-grid (red) geometries. The statistical measures (root mean squared deviation - RMSD, mean absolute deviation - MAD, and  $R^2$) for both cases are given in the figure and demonstrate the accuracy of the RKHS-PES.}\label{fig:FigureS4}
\end{figure}

The quality of the reproducing kernel Hilbert space (RKHS) interpolated PES (RKHS-PES) was assessed by comparing the interpolated energies with the corresponding \textit{ab initio} reference energies for both the on-grid and off-grid data sets. The agreement was quantified using the root mean square deviation (RMSD), mean absolute deviation (MAD), and $R^2$, as shown in Fig.~\ref{fig:FigureS4}.

\subsection{Collisional Dynamics}
\subsubsection{Quantum Dynamics}
Numerical parameters used in the quantum-dynamics (QD) calculations are listed in Table~\ref{tab:qd_parameters}.

\subsubsection{Cross sections}

The state-to-state cross sections for the quantum-dynamics (QD)
calculations were obtained using the partial wave summation according to
\begin{equation}
\sigma_{v_0j_0\Lambda_0, \beta v',j'}(E)
=
\frac{\pi}
{K_0^2(E)(2j_0+1)}
\mathcal{C}_{v_0j_0\Lambda_0,\beta v',j'}(E).
\label{eq:qd_cross_section}
\end{equation}
Here, \(K_0(E)=\sqrt{2\mu\left(E-E_{v_0j_0}\right)}/\hbar\) is the wave number associated with the initial scattering channel, $E$ is the energy, $\mu$ is the reduced mass of the collision system, $E_{v_0j_0}$ denotes the energy of the initial rovibrational eigenstate of the diatomic reactant and \(\mathcal{C}_{v_0j_0\Lambda_0,\beta v',j'}(E)\) represents a cumulative transition probability given by,
\begin{widetext}
\begin{equation}
\mathcal{C}_{v_0j_0\Lambda_0,\beta v',j'}(E)
=
\sum_J \sum_p \sum_{\Omega} \sum_{\Omega_0}
(2J+1)
\left|
S^{Jp}_{v_0j_0\Omega_0,\beta v',j'\Omega'}(E)
\right|^2.
\label{eq:cumulative_probability}
\end{equation}
\end{widetext}
Here, \(v_0\) and \(j_0\) denote the initial vibrational and rotational quantum numbers of the diatomic molecule, respectively, while \(v'\) and \(j'\) specify the corresponding quantum numbers after the collision. The quantities \(\Omega_0\) and \(\Omega'\) are the projections of the total angular momentum onto the body-fixed axis in the initial and final channels, respectively. The index \(\beta\) labels the asymptotic arrangement channel of the final scattering products, whereas \(\Lambda_0\) specifies the electronic state of the initial channel. Furthermore, \(J\) is the total angular momentum quantum number, \(p\) denotes the parity, and \(S^{Jp}_{v_0j_0\Omega_0,\,\beta v',j'\Omega'}(E)\) is the corresponding state-to-state element of the scattering \(S\)-matrix. The total elastic and inelastic cross sections were obtained by
summing the corresponding state-to-state cross sections over the
relevant final rovibrational states.

For the quasi-classical trajectory (QCT) calculations, the cross
section associated with an event \(x\) was evaluated as
\begin{equation}
\sigma_x
=
\pi b_{\max}^{\,2}P_x,
\label{eq:qct_cross_section}
\end{equation}
where \(b_{\max}\) is the maximum impact parameter and \(P_x\) is the
probability of the event. For an ensemble sampled uniformly in
\(b^2\), this probability is given by
\begin{equation}
P_x=\frac{N_x}{N_{\mathrm{tot}}},
\label{eq:qct_probability_simple}
\end{equation}
where \(N_x\) is the number of trajectories corresponding to event
\(x\), and \(N_{\mathrm{tot}}\) is the total number of trajectories.

In the present stratified-sampling calculations, the event
probability was instead evaluated using
\begin{equation}
P_x
=
\sum_{k=1}^{K}
V_k\frac{N_{x,k}}{N_k},
\label{eq:qct_probability_stratified}
\end{equation}
where \(N_{x,k}\) and \(N_k\) are, respectively, the number of
trajectories corresponding to event \(x\) and the total number of
trajectories in stratum \(k\), while \(V_k\) is the statistical weight
of that stratum satisfying
\(\sum_{k=1}^{K}V_k=1\).

The numerical values for the cross sections obtained are summarized in Table~\ref{tab:elastic_inelastic_cross_sections}.

\begin{figure}
    \centering
  \includegraphics[width=0.95\columnwidth]{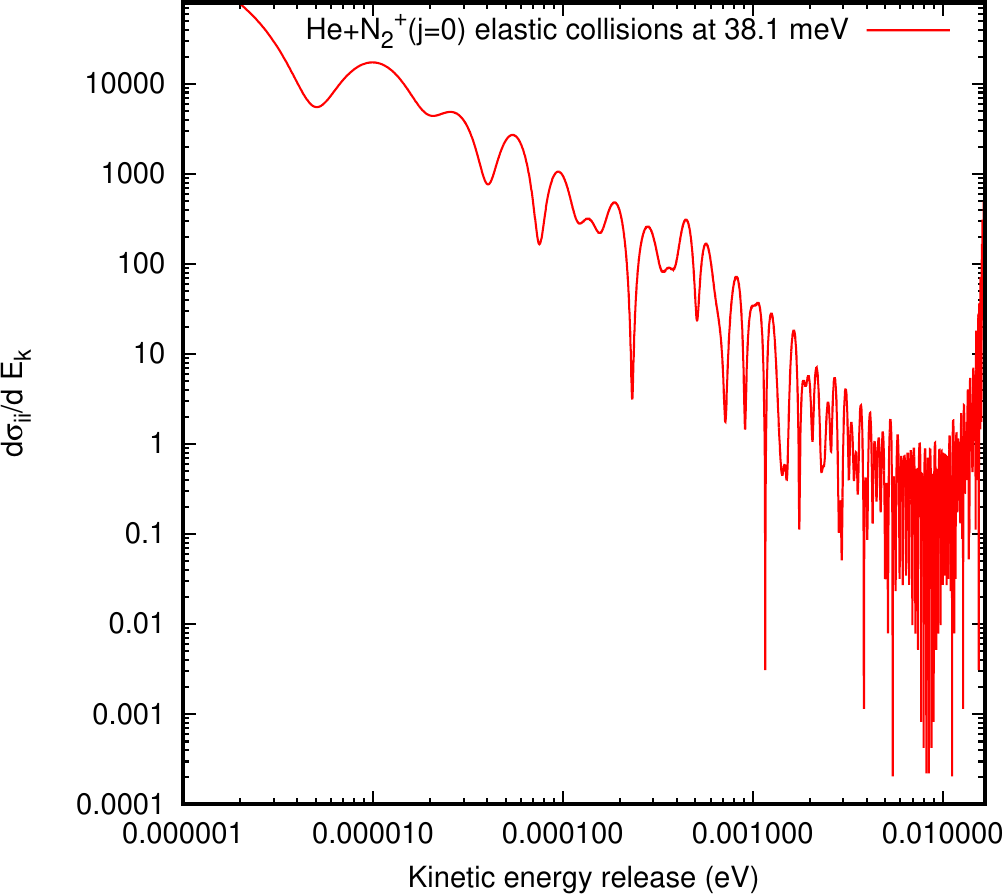}
  \caption{\textbf{Kinetic-energy transfer.} ${d\sigma_{ij}(E)\over d E_k}$ obtained with close-coupling time-independent calculations
    for He+N$_2^+(j=0)$ at 0.0381 eV, for total angular momenta $J=0,1,2,...,50$, including $\Omega=0,1,...\text{min}(10,J_{\text{tot}})$.
  The collision angle is described by 5000 equispaced points to obtain an accurate description of $E_k$ distribution.}
\end{figure}

\subsection{Energy transfer in quantum-dynamics calculations}

Experimentally, the elastic collisions of He atoms with trapped N$_2^+$ ions, initially at rest, require the transfer of sufficient energy to cause ion reordering, $\approx0.1$~meV in this case. 
The kinetic-energy transfer between He and N$_2^+$, $\Delta E_k$, depends on the deflection, $\theta$, defined as the angle between the initial velocity (parallel to the z-axis) and the final velocity in the centre-of-mass reference system.
Considering the conservation of energy and total momentum, $\Delta E_k$ can be defined as
\begin{eqnarray}\label{eq:energy-transfer-ions}
   \Delta E_k(\theta) =  {2 m_1  m_2\over (m_1+m_2)^2} E_1^0 (1-\cos\theta)
\end{eqnarray}
where $m_1$ and $m_2$ are the masses of He and N$_2^+$, respectively, and $E_1^0=38.1$ meV
is the initial kinetic energy of the He atoms with N$_2^+$ is assumed to be at rest.

Quantum mechanically, the scattering angle $\theta$ is described by a distribution given by the differential cross section
\begin{widetext}
\begin{eqnarray}\label{difXsect}
  {d\sigma_{ij}(E)\over d \Omega} = {1\over 4 k_i^2}  \left\vert \sum_J (2J+1) \left\lbrace S^{J}_{ij}(E) -\delta_{ij}\right\rbrace
    d^J_{\Omega,\Omega'}(\theta)\right\vert^2,
\end{eqnarray}
\end{widetext}
for a collision  of an atom with a diatomic molecule in a body-fixed frame (whose $z$-axis is defined along the vector between
the centre-of-mass of He and N$_2^+$, {\bf R}, with the three atoms in the $x-z$ body-fixed plane), 
where $d^J_{\Omega,\Omega'}(\theta)$ are Wigner matrices with $\Omega,\Omega'$ being the projections of $J$ and $j$ (total 
and diatomic rotation angular momenta) in the body-fixed $z$-axis for the initial, $i$, and final, $j$, states.

Using Eq.~(\ref{eq:energy-transfer-ions}), the distribution  associated to the  kinetic energy  transferred to N$_2^+$, $\Delta E_k$, is given by 
\begin{eqnarray}\label{EdifXsect}
  {  d \sigma_{i,j} \over d E_k } & =&   {  d \sigma_{i,j} \over d \Omega }  {  d \Omega \over d E_k } 
\end{eqnarray}

For the elastic collision for $j,j'=0$, $d^J_{\Omega,\Omega'}(\theta)$ becomes $P_J(\theta)$ and the kinetic energy gained by N$_2^+$ ($m_2$)
becomes
\begin{widetext}
\begin{eqnarray}
  {d\sigma_{ii}(E)\over d E_k} = {\pi\over 4 k_i^2} {(m_1+m_2)^2\over m_1m_2}  E_1^0\left\vert \sum_J (2J+1) \left\lbrace S^{J}_{ii}(E) -\delta_{ij}\right\rbrace
    P_J(\theta)\right\vert^2,
\end{eqnarray}
which is the kinetic energy distribution, shown in the figure.
\end{widetext}

The fraction of effective elastic collision is obtained as $f_{>}/f_{\text{tot}}$= 0.57,
where $f_\text{{tot}}$ is the integral of the distribution in the whole kinetic energy transferred,
while $f_{>}$ is the integral for $E_k> 0.1$ meV.

\section{Evaluation of experimental uncertainties}

\subsection{Uncertainties in observation of collision events}\label{sec:unc-in-collsion-event-observation}

\subsubsection{Background collisions}
In principle, collisions with background-gas molecules could be incorrectly attributed to collisions with the molecular beam. To assess the background collision probability, collision markers for the Ca$^+$-N$_2^+$ string were tracked for a total of 3710~s without exposure to the molecular beam. No fluorescence losses and 3 ion reorderings were observed in these measurements yielding a background reordering rate of $\sim8 \times10^{-4}$~s$^{-1}$. This result is more than two orders of magnitude lower than the molecular-beam-induced ion-position change rate for the N$_2$/He mixed beam which was estimated at 0.19(1)~s$^{-1}$. The probability of misidentification of background collisions with the collisions induced by the molecular beam at rate $\lambda_{\mathrm{MB}}$, is thus
$P(\mathrm{bcg}\mid \mathrm{event}) =
\lambda_{\mathrm{bcg}} / (\lambda_{\mathrm{MB}} + \lambda_{\mathrm{bcg}}) \, ,$
which yields a probability of approximately 0.4\%. 

Assuming reordering rates record only 50\% of collisions, the actual background scattering rate was twice as high, i.e. $\lambda_{\mathrm{bcg}}^{\mathrm{tot}}$ = 2$\lambda_{\mathrm{bcg}}$ $\approx$ 1.6 $\times$ 10$^{-3}$ s$^{-1}$ per two ions. From this collision rate, one can estimate the pressure near the trap region to be around 2 $\times$ 10$^{-11}$ mbar which is compatible with the reading on the pressure gauge in the experiment \cite{pagano2018cryogenic-b58}.  

In this analysis, the pulsed nature of the molecular beam was ignored, i.e., that collisions should occur solely during the time windows when the molecular pulses pass through the trap centre. In principle, measuring the exact timestamps of the collision events and correlating them with the passage of the molecular beam would allow a high-fidelity discrimination of almost all background events. However, in the present experiments, the identification of reordering events was limited by the temporal resolution of the camera preventing the exact determination of the time of collision events.

\subsubsection{Probability of multiple collisions}

Depending on the amount of kinetic energy deposited in the ion string and the efficiency of sympathetic cooling, features in fluorescence traces or changes in ion position on the camera could persist for many hundreds of ms, posing a limit on the distinguishability of two consecutive collision events. Also, the temporal resolution of the camera restricted observation of rapid ion reorderings in the trap. In the data analysis, a minimum time difference of at least 0.6 s between collision observables was assumed to assign them to two separate collisions events. Based on the measured collision rates discussed above, the probability of two or more unresolved collisions in one collision attempt was thus estimated to be $\sim$ 3\%.

\subsubsection{Uncertainties of collision numbers} \label{sec:uncertainties-collision-number}

The expected number of collisions $n_i$ for each scattering pair $i$ was calculated as outlined in Sec. \ref{sec:collision-partner-assignment}. The total variance of $n_i$ can be approximated by a sum of independent contributions from multinomial sampling statistics, the uncertainty in the total number of collisions and uncertainty of the collision probabilities:
\begin{equation}
\sigma_{n_i}^2
= \underbrace{N p_i (1 - p_i)}_{\text{multinomial}}
+ \underbrace{(\delta_N^{\mathrm{rel}})^2 (N p_i)^2}_{\#\text{collisions}}
+ \underbrace{N^2 \,\mathrm{Var}(p_i)}_{\text{probability unc.}}
\end{equation}
with total number of collisions, $N$, having relative uncertainty of $\delta_N^{\mathrm{rel}}$, and probabilities $p_i$ of each partner pair participating in a collision. 

\subsection{Uncertainties in observing N$_2^+$ state-changing collisions}\label{sec:uncertainty-of-N2-state-change}

The relative uncertainty in observing a state-change due to collisions with a molecular beam was estimated to be $\sim$~1\%. Below, we discuss particular contributions to this error arising from background scattering and experimental factors.

\subsubsection{State-changing background collisions}
A typical single iteration of a collision experiment including the exposure to the molecular beam and two series of state-detection pulses, took approximately 48 s. During this time, there was a small probability that background collisions caused rotational excitation of the molecule that was not related to a collision with gas from the molecular beam.

From the background collision rate, $\lambda_{\mathrm{bcg}}^{\mathrm{tot}}$, we infer the background collision probability per collision attempt cycle (48~s) to be 3.8\% per ion (Sec.~\ref{sec:unc-in-collsion-event-observation}). However, only a small fraction of background collisions would be inelastic, with most of the collisions being either elastic or reactive. The latter is due to collisions with neutral H$_2$, a dominant vacuum contaminant, or with H$_2$O. Such reactive collisions were clearly identified by the mass change of the trapped dark ion and were omitted in the data analysis.

Conservatively assuming that 10\% of background collisions were inelastic, the probability of a state change due to background collisions during a single experimental cycle would be below 0.5\%.

\begin{figure*}
    \includegraphics[width=0.85\linewidth]{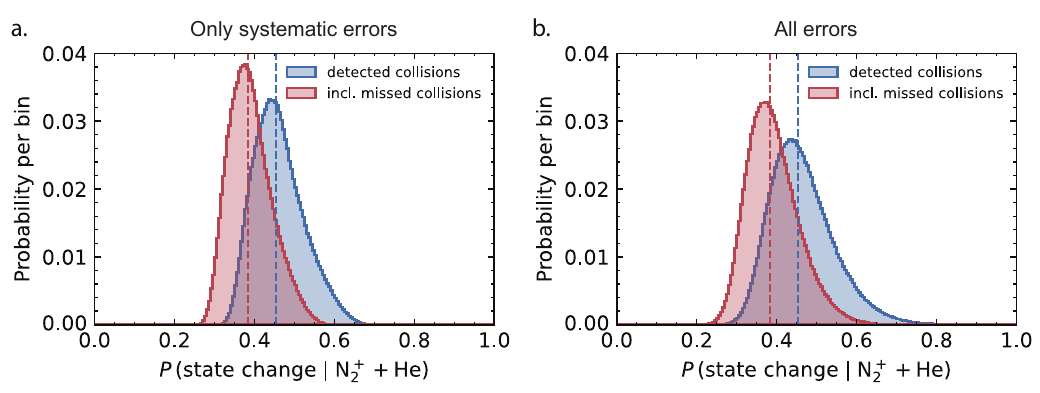}
    \caption{\textbf{Distribution of state-change probabilities obtained from Monte Carlo simulations} when accounting for systematic shifts only (a) and when including all errors (b). The dashed lines are the nominal values obtained from the experiments without accounting for the systematic shifts. The bin width was set to 0.005.}
    \label{fig:FigureS7}
\end{figure*}

\subsubsection{Experimental factors}
The statistical fidelity of our state detection was previously estimated to exceed 99.9\% \cite{sinhal2020quantum-nondemolition-757, roguski2025role-43b}. However, environmental factors such as background collisions or cooling imperfections could affect it. By performing two sets of state-detection pulses separated by an additional time for sympathetic cooling, such issues were mitigated. Additionally, in the post-processing of the data, instances of rotational state changes from off-resonant scattering caused by the lasers used for state detection were identified excluded from the data set.

Another small error contribution may stem from the fact that during data treatment, we included a single state-changing event associated with both a reordering and a decrystallisation event and assigned the state change to only one of them. 

Overall, we anticipate that the state-detection uncertainty should not exceed 0.5\%.

\subsection{Estimation of systematic errors due to assumed collision model} \label{sec:systematic-errors-in-data-analysis}

The calculation of the expected number of collisions for each ion-neutral pair in the experiments was based on the relative Langevin rates corrected for the composition of the molecular beam (see Sec.~\ref{sec:collision-partner-assignment}). To evaluate the effect of deviations of the individual collision cross sections from the Langevin model on the final inferred inelastic N$_2^+$-He scattering probability, a sensitivity analysis was performed. Conservatively assuming relative deviations of up to 30\% of each individual collision rate from the Langevin values, Monte Carlo sampling of all rates was performed across the assumed uncertainty interval with a sample size of 10$^6$. Additionally, in this analysis the measured inelastic scattering probability of the N$_2^+$+N$_2$ collisions was allowed to vary by up to $\pm$ 30\%. This step accounted for possible deviations in the observed inelastic scattering probability due to differences in collision conditions, such as collision energies for this pair, which were significantly higher in the mixed-beam experiments than in those using a pure N$_2$ beam. Fig.~\ref{fig:FigureS7}a shows the resulting probability distributions. The inferred mean state-change probability is reported in Table~\ref{tab:pHe_uncertainty}.

The analysis was further extended to include the statistical uncertainties summarised in Secs.~\ref{sec:uncertainties-collision-number} and~\ref{sec:uncertainty-of-N2-state-change}, i.e. due to multinomial sampling, uncertainty in the total collision number, uncertainty of mixing ratio preparation and state-detection uncertainty. The latter three were evaluated with Gaussian sampling to account for their stochastic origin. The resulting distribution of the inelastic N$_2^+$-He scattering probability is shown in Fig.~\ref{fig:FigureS7}b.


\section{Determination of collision energies} \label{sec:collision-energies}

The initial internal energies of the ionic collision partners were negligible in the present experiments. While the trapped N$_2^+$ ions were prepared in their rotational ground state, Ca$^+$ was continuously Doppler laser cooled during the collision experiments. The contribution to the scattering cross-section of Ca$^+$ from electronically excited states was neglected \cite{knoop1995lifetime-7b0, moriwaki2000laser-67b}. The supersonic molecular beam was internally cold, with He in its electronic ground state and N$_2$ in its vibrational ground state confined to the few lowest rotational states \cite{Tong2010}. 

The collision energies were dominated by the kinetic energy of the molecular beam whose velocities were estimated using the expression for its terminal velocity after a supersonic expansion of the gas \cite{wall2016preparation-f46},
\begin{equation}\label{eq:supersonic-beam-velocity}
   v_{\mathrm{beam}} =\sqrt{\frac{2\gamma_{\mathrm{eff}}}{\gamma_{\mathrm{eff}} - 1}
   \frac{RT_0}{M_{\mathrm{eff}}}}.
\end{equation}
$M_{\mathrm{eff}}$ is a sum of molar masses of the component species ($M_{\rm{He}}$, $M_{\rm{N_2}}$) scaled by the mole fraction ($x_{\rm{He}}$, $x_{\rm{N_2}}$) of each component, i.e. $M_{\mathrm{eff}} = x_{\rm{He}} M_{\rm{He}} + x_{\rm{N_2}} M_{\rm{N_2}}$. $\gamma_{\mathrm{eff}}$ is the ratio of the specific heat capacities of the mixture, also including contributions from each beam component. $R$ is the ideal gas constant and $T_0$ is the stagnation temperature.  The estimated velocities for the beam containing He and N$_2$ with a ratio of 7:1 and for a beam containing pure N$_2$ as used in the control experiments are listed in Table \ref{tab:collisional-energies}. 

The collision energies $E_\text{CM}$ for different collision pairs are also listed in Table \ref{tab:collisional-energies}. The trapped ions were continuously Doppler laser cooled to mK temperatures and can be considered effectively stationary in the lab frame. The instantaneous secular and micromotion energies associated with the ions movement in an RF trap were of order $\mu$eV and thus negligible. 


\newpage
\section{Additional theoretical results}
\label{sec:additional_theory}

\begin{figure}[h!]
\centering
\includegraphics[width=0.9\columnwidth]{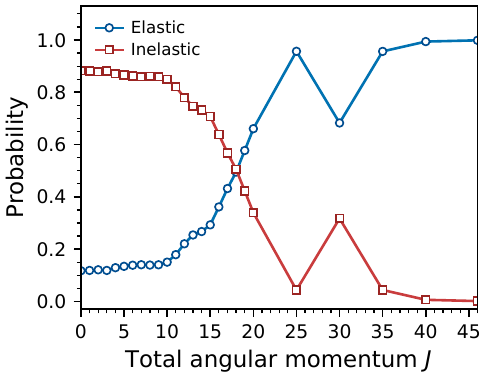}
\caption{\textbf{QD elastic and inelastic probabilities} as a function of the total-angular-momentum quantum number $J$.}
 \label{fig:FigureS5}
\end{figure}

\begin{figure}[h!]
\centering
\includegraphics[width=0.9\columnwidth]{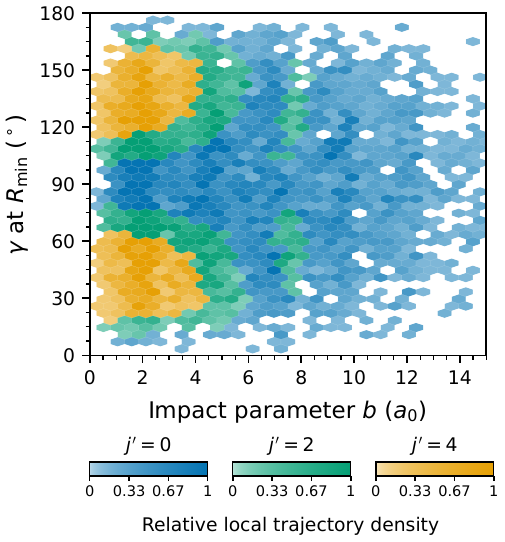}
\caption{
\textbf{Final-state-resolved closest-approach dynamics.}
Distribution of the Jacobi angle $\gamma$ at the minimum intermolecular separation of the scattering trajectories, $R_{\min}$, as a function of the impact parameter $b$. Each hexagonal bin is coloured according to the dominant final rotational state, $j'=0$, 2 or 4, among the trajectories contained within that bin. The colour intensity reflects the probability. }
 \label{fig:FigureS6}
\end{figure}


\pagebreak\clearpage
\onecolumngrid 

\section*{Tables}


\begin{table*}[!h]
\caption{\textbf{Collision detection statistics.} The probability of a state change of N$_2^+$ (fourth column) for each collision marker is calculated from the number of state changes (third column) per number of observed events (second column). For the one state-changing event accompanied by both reordering and decrystallisation, it was assumed that only one of these events caused the state change. Note that this analysis pertains to the raw data of observed collisions before considering missed events (Sec.~\ref{sec:undetected-collisions}) and the role of the gas mixture in the beam (Sec.~\ref{sec:collision-partner-assignment}).}
\label{tab:appendix-col-det-stat}

\begin{tabular}{@{}llll@{}}
\toprule
 & \# of events & N$_2^+$ state changes & N$_2^+$ state-change \\
& & & probability \\
\midrule
Reordering (no decrystallisation)    & 136   & 21  & 15\%  \\
Decrystallisation (no reordering)  & 38   & 21  & 55\%  \\
Reordering + decrystallisation & 3+3   & 1  & --  \\
Fluorescence loss w/o other markers    & 7   & 0  & --  \\
Fluorescence loss with other markers    & 42   & 14  & 33\%  \\
Total num. obs. collisions    & 187   & 43  & 23\%  \\
\botrule
\end{tabular}
\end{table*}


\begin{table*}[!h]
\centering
\caption{\textbf{Collision probabilities $P$ for possible pairs of collision partners} assuming Langevin ($k_L$, model I) and semiclassical ($k_{\mathrm{SC}}$, model II) collision rate coefficients. See text for details.}\label{tab:langevin-rates-probability}
\begin{tabular}{lllll c}
Ion & Neutral & $k_\mathrm{L}$ ($\times$ 10$^{-16}$ m$^3$/s) & $k_{\mathrm{SC}}$ ($\times$ 10$^{-15}$ m$^3$/s) & $P$ (model I) & $P$ (model II)\\
\hline
Ca$^+$ & He & 5.6 & 4.15 & 0.412(9) & 0.31(1) \\
Ca$^+$ & N$_2$ & 7.7 & 17.4 & 0.081(9) & 0.19(1)\\
N$_2^+$ & He & 5.7 & 4.15 & 0.420(9) & 0.31(1)\\
N$_2^+$ & N$_2$ & 8.3 & 17.4 & 0.088(9) & 0.19(1)\\
\hline
\end{tabular}
\end{table*}


\begin{table*}[!h]
\caption{\textbf{Collision-detection statistics for the experiments with a molecular beam containing only N$_2$.}  Note that this analysis pertains to the raw data of observed collisions before considering missed events. }\label{tab:col-det-stat-pureN2}
\begin{tabular}{@{}llll@{}}
\toprule
 & Number  & Number of  & Probability  \\
 & of events & state-changing events & of state change \\ 
\midrule
Reordering (no decrystallisation)    & 78   & 14  & 18\%  \\
Decrystallisation (no reordering)  & 11   & 7  & 64\%  \\
Reordering + Decrystallisation (same attempt) & 0   & -  & -  \\
Fluorescence loss w/o other markers    & 5   & 1  & 20\%  \\
Fluorescence loss with other markers    & 10   & 5  & 50\%  \\
Total num. obs. collisions    & 94   & 22  & 23\%  \\
\botrule
\end{tabular}
\end{table*}


\begin{table*}[!h]
    \centering
    \caption{\textbf{Expected number of collision events.} The values were calculated for the two collision models discussed in Sec. \ref{sec:collision-partner-assignment}, and rounded to the nearest integer. The table contains the numerical data displayed in Fig.~\ref{fig:fig3-collision-statistics} of the main text. Due to rounding to integer values in the table, the reported individual collision counts may not sum exactly to the reported total number of collisions. The inelastic scattering probabilities were calculated using the unrounded values.}
    \label{tab:appendix-raw-data-figure-statistics}
    \begin{tabular}{lcccc}
    \hline
    & \multicolumn{2}{c}{Model I}
    & \multicolumn{2}{c}{Model II} \\
    \hline
     & Observed & Incl. missed & Observed & Incl. missed\\
    \hline
    Total number of collisions & $187 \pm 6$ & $281 \pm 9$ \\
    Ca$^+$ + N$_2$ & $15 \pm 4$ & $23 \pm 5$ & $35 \pm 6$ & $53 \pm 8$ \\
    N$_2^+$ + N$_2$ & $16 \pm 4$ & $25 \pm 5$ & $35 \pm 6$ & $53 \pm 8$ \\
    Ca$^+$ + He & $77 \pm 7$ & $116 \pm 9$ & $58 \pm 7$ & $88 \pm 9$ \\
    N$_2^+$ + He & $78 \pm 7$ & $118 \pm 9$ & $58 \pm 7$ & $88 \pm 9$ \\
    N$_2^+$ + N$_2$ state-change & $7 \pm 2$ & $11 \pm 3$ & $16 \pm 3$ & $24 \pm 4$ \\
    N$_2^+$ + He state-change & $36 \pm 2$ & $45 \pm 3$ & $27 \pm 3$ & $32 \pm 4$ \\
    \hline
    N$_2^+$ + He inelastic probability & $(45\pm 5)$\% & $(38 \pm 4)$\% & $(46 \pm 8)$\% & $(37 \pm 6)$\% 
    \end{tabular}
\end{table*}


\begin{table*}[!h]
\centering
\caption{\textbf{Numerical parameters used in the quantum-dynamics (QD) calculations.}
The table summarizes the coordinate ranges, grid sizes,
absorbing potential parameters, number of propagation iterations,
and maximum helicity quantum number.} \label{tab:qd_parameters}
\begin{tabular}{lc}
\hline
\textbf{Parameter} & \textbf{Value} \\
\hline
$r_{\min}, r_{\max}, n_r$
    & $0.8,\ 1.6\ \mathrm{\AA},\ 72$ \\

$R_{\min}, R_{\max}, n_R$
    & $0.9,\ 27\ \mathrm{\AA},\ 384$ \\

$n_{\gamma}$
    & $80$ \\

$R_{\mathrm{abs}}, \alpha_{\mathrm{abs}}$
    & $15\ \mathrm{\AA},\ 10^{-6}\ \mathrm{\AA}^{-6}$ \\

$n_{\mathrm{iterations}}$
    & $12500$ \\

$\Omega_{\max}$
    & $\min(J,19)$ \\
\hline
\end{tabular}
\end{table*}


\begin{table*}[htbp]
\centering
\centering
\caption{\textbf{Elastic, inelastic, and total cross sections for  N$_2^+$+He collisions.} Cross sections obtained from quantum dynamics (QD) and quasi-classical trajectory (QCT) calculations are compared with the previously reported QCT inelastic cross section of Paláček \textit{et al.}~\cite{Palacek2024} who used a PES at the lower MCSCF level of theory and a modified aug-cc-pVQZ basis set. All cross sections are reported in \AA$^2$.} \label{tab:elastic_inelastic_cross_sections} 
\begin{tabular}{lccc} 
\toprule 
Method & $\sigma_{\mathrm{elastic}}$ (\AA$^2$) & $\sigma_{\mathrm{inelastic}}$ (\AA$^2$) & $\sigma_{\mathrm{total}}$ (\AA$^2$) \\ \midrule 
QD & 92.6 & 24.4 & 117.0 \\ 
QCT & 168.4 & 29.5& 197.9 \\ 
QCT (Ref.~\cite{Palacek2024})
    & --- & 21.0 & --- \\
\bottomrule 
\end{tabular}
\end{table*}


\begin{table*}[!h]
\centering
\caption{{\bf State-change probability for $\mathrm{N_2^+ + He}$ obtained from Monte Carlo simulations.}
Uncertainties correspond to the standard deviation of the
distributions shown in Fig.~\ref{fig:FigureS7}, and similarly when including only statistical errors. The nominal value was obtained without accounting for possible systematic shifts as discussed in Sec.~\ref{sec:extracting-ratio-of-elastic-to-inelastic}.}
\label{tab:pHe_uncertainty}
\begin{tabular}{lcccc}
\hline
Dataset & Nominal & Only syst. errors & Only stat. errors & All errors \\
\hline
Detected
& $0.45 \pm 0.05$
& $0.46 \pm 0.06$
& $0.46 \pm 0.04$
& $0.46 \pm 0.08$ \\

Including missed
& $0.38 \pm 0.04$
& $0.39 \pm 0.05$
& $0.39 \pm 0.03$
& $0.39 \pm 0.06$ \\
\hline
\end{tabular}
\end{table*}


\begin{table*}[!h]
\centering
\caption{\textbf{Beam velocities $v_\text{beam}$ and centre-of-mass collision energies $E_\text{CM}$} calculated for a reservoir temperature of 295 K. "Seeded beam" refers to a beam with a mixing ratio of 7:1 He:N$_2$ used in the experiments. Values for collisions with a pure N$_2$ beam are included for comparison with the control measurements discussed in Sec.~\ref{sec:appendix-experiemnts-with-pure-N2-beam}.}\label{tab:collisional-energies}
\begin{tabular}{lcc}
\hline
\textbf{Collision partners} & \textbf{$v_{\mathrm{beam}}$ (m/s)} & \textbf{$E_{\mathrm{CM}}$ (meV)} \\
\hline
N$_2$ (seeded beam) + Ca$^+$ & 1356 & 157 \\
N$_2$ (seeded beam) + N$_2^+$ & 1356 & 133 \\
He (seeded beam) + Ca$^+$   & 1356 & 35 \\
He (seeded beam) + N$_2^+$  & 1356 & 33 \\
\\
N$_2$ (pure beam) + N$_2^+$ & 783  & 44 \\
N$_2$ (pure beam) + Ca$^+$  & 783  & 52 \\
\hline
\end{tabular}
\end{table*}

\end{document}